\documentclass[aps,pre,twocolumn,showpacs,10pt]{revtex4}
\usepackage{color}
\usepackage{amsmath}
\usepackage{amsfonts}
\usepackage{amssymb}
\usepackage{epsfig}
\usepackage{graphicx}

\newcommand{\vib}{\mbox{\scriptsize vib}}
\newcommand{\eSP}{e_{\mbox{\scriptsize SP}}}
\newcommand{\eIS}{e_{\mbox{\scriptsize IS}}}
\newcommand{\eTH}{e_{\mbox{\scriptsize th}}}

\newcommand{\fit}{{\mbox{\scriptsize (fit)}}}

\begin{document}

\title{Equilibrium phase diagram of a randomly pinned glass-former}

\author{Misaki Ozawa}
\affiliation{Institute of Physics, University of
Tsukuba, Tsukuba \mbox{305-8571,~Japan} and Department of Physics, Nagoya
University, Nagoya 464-8602, Japan}

\author{Walter Kob}
\affiliation{Laboratoire Charles Coulomb, UMR 5221,
University of Montpellier and CNRS, Montpellier, France}

\author{Atsushi Ikeda}
\affiliation{Fukui Institute for Fundamental Chemistry, Kyoto
University, Kyoto, 606-8103, Japan}

\author{Kunimasa Miyazaki}
\affiliation{Department of Physics, Nagoya
University, Nagoya 464-8602, Japan}

\begin{abstract}
We use computer simulations to study the thermodynamic properties
of a glass former in which a fraction $c$ of the particles has been
permanently frozen. By thermodynamic integration, we determine the
Kauzmann, or ideal glass transition, temperature $T_K(c)$ at which the
configurational entropy vanishes. This is done without resorting to any
kind of extrapolation, {\it i.e.}, $T_K(c)$ is indeed an equilibrium
property of the system. We also measure the distribution function of the
overlap, {\it i.e.}, the order parameter that signals the glass state.
We find that the transition line obtained from the overlap coincides
with that obtained from the thermodynamic integration, thus showing
that the two approaches give the same transition line. Finally we
determine the geometrical properties of the potential energy landscape,
notably the $T-$and $c-$dependence of the saddle index and use these
properties to obtain the dynamic transition temperature $T_d(c)$. The
two temperatures $T_K(c)$ and $T_d(c)$ cross at a finite value of $c$
and indicate the point at which the glass transition line ends. These
findings are qualitatively consistent with the scenario proposed by the
random first order transition theory.
\end{abstract}

\keywords{glass materials | amorphous order |ideal glass transition}

\maketitle

Upon cooling, glass-forming liquids show a dramatic increase of
their viscosities and relaxation times before they eventually fall out of
equilibrium at low temperatures~\cite{debenedetti2001,Binder2011}. This
laboratory glass transition is a purely kinetic effect since it occurs at
the temperature at which the relaxation time of the system crosses the
time scale imposed by the experiment, {\it e.g.}, via the cooling rate.
Despite the intensive theoretical, numerical, and experimental studies of
the last five decades, the mechanism responsible for the slowing down and
thus for the (kinetic) glass transition is still under debate and hence a
topic of intense research. From a fundamental point of view the ultimate
goal of these studies is to find an answer to the big question in the
field: Does there exist a {\it finite} temperature at which the dynamics
truly freezes and, if it does, whether this {\it ideal} glass transition
is associated with a thermodynamic singularity or whether it is of kinetic
origin~\cite{Cavagna2009b,Berthier2011j,Berthier2011rmp,Chandler2010arpc}.

Support for the existence of a kinetic transition comes from
certain lattice gas models with a ``facilitated dynamics''
\cite{Chandler2010arpc}.  In these models, the dynamics is due to
the presence of ``defects'' and hence for such systems the freezing
is not related to any thermodynamic singularity.  However, the first
evidence that there does indeed exist a thermodynamic singularity
goes already back to Kauzmann who found that the residual entropy
(the difference of the entropy of the liquid state from that of
the crystalline state) vanishes at a finite temperature $T_K$
if it is extrapolated to temperatures below the laboratory glass
transition~\cite{Kauzmann1948}. Subsequently many theoretical scenarios
that invoke the presence of a thermodynamic
transition have been proposed~\cite{Tarjus2010,Tanaka2011,Biroli2012wolynesbook}. 
One of these is the so-called ``random first order transition'' (RFOT)
theory which, inspired by the exact solution of a mean-field spin
glass, predicts that at $T_K$ the glass-former does indeed undergo
a thermodynamic transition at which the residual, or configurational
entropy $S_{\rm c}$ (the logarithm of the number of the states which
are available to the system) vanishes and concomitantly breaks the
replica symmetry~\cite{kirkpatrick1989,Biroli2012wolynesbook}. A further
appealing feature of RFOT is that it seems to reconcile in a natural way
the (free) energy-landscape scenario and mode-coupling theory (MCT),
a highly successful theory that describes the relaxation dynamics at
intermediate temperatures~\cite{Gotze2009}.

Despite all these advances, the arguments put forward in the various
papers must be considered as phenomenological since compelling and
undisputed experimental or numerical evidence to prove or disprove any of
these theories and scenarios is still lacking. The only exception
are hard spheres in infinite dimensions, for which mean-field theory
should become exact~\cite{Kurchan2013jpcb}, but even in this case some
unexpected problems are present, see Ref.~\cite{Ikeda2010prl}. This lack
of understanding is mainly due to the steep increase of the relaxation
times which hampeirs the access to the transition point of thermally
equilibrated systems and hence most of the efforts to identifying the
transition point, if it exists, resort on unreliable extrapolation.

\section{Randomly pinned systems}

Recently a novel idea to bypass this difficulty has been
proposed~\cite{Cammarota2012pnas,Cammarota2013jcp,Cammarota2013epl,Berthier2012pre}.
By freezing, or pinning, a fraction of the degrees of freedom of the
system, the ideal glass transition temperature has been predicted to rise
to a point at which experiments and simulations {\it in equilibrium}
are feasible thus allowing to probe the nature of this transition.
In  Ref.~\cite{Cammarota2012pnas} the authors have studied the effect
of pinning for the case of a mean-field spin glass model which is
known to exhibit a dynamical MCT-transition at a temperature $T_d$
and a thermodynamic transition at a lower temperature $T_K$.  It was
demonstrated that, by pinning a fraction $c$ of the degree of freedoms 
of the spins (selected at random) in the equilibrated system,
both $T_K(c)$ and $T_d(c)$ increase with $c$. Thus by equilibrating
the non-pinned system at an intermediate temperature and subsequently
increasing $c$, one can access and probe the ideal glass state in
{\it thermal equilibrium} (Note that changing $c$ does not perturb this
equilibrium~\cite{Scheidler2004jpcb}).  It was found that at sufficiently
large $c$ the two lines $T_K(c)$ and $T_d(c)$ merge and terminate at
a critical point with a universality class of the random-field Ising
model~\cite{Cammarota2012pnas,Franz2013jsmte3}.

Very recently, it has been tested whether this approach to detect $T_K$
in mean field models can also be used in realistic glass formers in
finite dimensions~\cite{Kob2013prl}. It was found that pinned systems do
indeed show a behavior that agrees qualitatively with the theoretical
predictions, thus giving encouraging evidence that the nature of the
ideal glass transition can be studied in equilibrium. In these numerical
studies, the overlap $q$ and its distribution $[P(q)]$ have
been used to identify the amorphous-fluid phase diagram in the $T$-$c$
plane for pinned systems~\cite{Kob2013prl}.

Despite these results, it is not clear if the so obtained amorphous
state is the {\it bona fide} ideal glass, for which the configurational
entropy $S_{\rm c}$ vanishes and, in view of the conceptual importance
of $S_{\rm c}$, this is a very disturbing situation.  We recall
that $S_{\rm c}$ is related to the number of available states of the
system~\cite{kirkpatrick1989,Biroli2012wolynesbook} and is also a key
quantity that controls the slow dynamics in the activation regime
in which the relaxation time $\tau_\alpha$ is related to $S_{\rm
c}$ via the Adam-Gibbs relation, $\ln \tau_\alpha \propto 1/TS_{\rm
c}$~\cite{adam1965,Biroli2012wolynesbook}.  Finally, $S_{\rm c}$ is
intimately related to the Landau-like free energy associated to the
overlap of two coupled replicas~\cite{Franz1997}.

Although from the simulations reported in Ref.~\cite{Kob2013prl}, the
existence of $T_K$ has been inferred from the behavior of the overlap
distribution $[P(q)]$ (discussed in more detail below),
such an analysis of the overlap $q$ alone may not be conclusive to
demonstrate the existence of the arrested phase, since one can not exclude
the possibility that other scenarios, such as the purely kinetic one,
can also explain the observed features of $[P(q)]$.  Thus,
evaluating $S_{\rm c}$ directly and identifying the temperature at which
it vanishes is crucial to disentangle conflicting theoretical scenarios.

In the present work, we use computer simulation to determine for a
canonical glass-former with pinned particles the ideal glass transition
temperature $T_K(c)$ as a point at which $S_{\rm c}$ vanishes. For the
first time this is done without invoking any kind of extrapolation. We
also calculate the overlap distribution $[P(q)]$ and
find that $T_K$ obtained from $S_{\rm c}(T_K)=0$ and from $[P(q)]$
agree very well up to a finite value of $c$. Furthermore, we analyze
the geometrical properties of the potential energy landscape (PEL)
and use it to evaluate the dynamic transition temperature $T_d(c)$ as
a point at which the saddles of the energy landscape vanish.  We find
that $T_d(c)$ merges with $T_K(c)$ exactly at the point at which the
two mentioned $T_K(c)$ depart from each other, strongly indicating
the existence of a critical point which is predicted by the mean-field
analysis~\cite{Cammarota2012pnas,Cammarota2013jcp,Cammarota2013epl}.

\section{Results}
We study a standard glass-forming model: A three dimensional binary
Lennard-Jones mixture~\cite{Kob1995c}.  The number of particles is
$N=150$ and 300, but most of the results are for $N=300$ (see {\it
Materials and Methods} for details). Throughout the present study,
the system has been prepared at each temperature by randomly choosing
a fraction $c$ of particles from the thermally equilibrated samples and
quenching their positions  (see {\it Materials and Methods} for details).

\subsection{Entropy and configurational entropy}

To obtain the entropy of the pinned system, $S$, we used thermodynamic
integration to determine the entropy of a given configuration of
pinned particles and subsequently calculated $S$ by averaging over
the realizations of pinned particles (see {\it Materials and Methods}
for details).

Figure~1(a) shows the entropy per (unpinned) particle $s\equiv S/N(1-c)$
as a function of the fraction of pinned particles at several temperatures
$T$ and we recognize that with increasing $c$ the entropy decreases
rapidly. For all temperatures this decrease is linear at small $c$
but then the curves bend at intermediate $c$ and follow a weaker
$c$-dependence. For $T\lesssim 0.5$ this bent becomes sharp,
strongly indicating that a {\sl thermodynamic} glass transition takes
place.

\begin{figure}[th]
\begin{center}
\includegraphics[width=0.48\textwidth,clip]{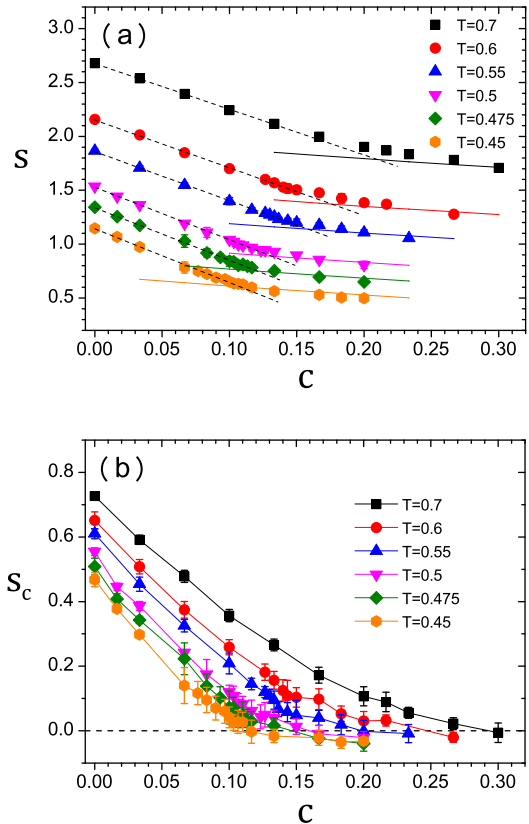}
\caption{
(a): Entropy of the system, $s$, as evaluated from the thermodynamic
integration, as a function of $c$ (symbols). The entropies of the
disordered solid states $s_{\rm vib}$, obtained using the harmonic
approximation, are drawn as solid lines.  The dashed lines are a linear
extrapolation from the low $c$ sides. (b): The configurational entropy
$s_{\rm c}=s-s_{\rm vib}$. The error bars have been estimated from the
sample to sample fluctuations.
}
\end{center}
\label{fig1}
\end{figure} 
 
This becomes more evident by evaluating the configurational entropy obtained
by subtracting from $S$ the vibrational entropy $S_{\rm vib}$.
In order to estimate $S_{\rm vib}$, we have determined the inherent
structures~\cite{Stillinger1982} and calculated the eigenfrequencies
$\omega_a$. Using the harmonic approximation, one can then approximate
$S_{\rm vib}=\sum_a \{ 1- \log(\beta \hbar \omega_a)\} $.  $s_{\rm
vib}\equiv S_{\rm vib}/N(1-c)$ is shown in Fig.~1(a) as well (solid
lines) and we see that it shows basically a linear decrease with $c$,
a trend which is due to the suppression of the low frequency modes in
the density of states.

We can now estimate the configurational entropy $S_{{\rm c}}$ as
the difference $S_{\rm c}=S-S_{\rm vib}$~\cite{Sciortino1999prl} and
in Fig.~1(b) we show the $c$-dependence of $s_{\rm c}= S_{\rm c}/N(1-c)$
for various temperatures. This figure shows that, for $T \lesssim 0.5$, $s_{\rm c}$
quickly decreases with increasing $c$ and becomes basically zero {\it
at a finite value of $c$}, indicating that the system has entered
the ideal glass state in which the entropy is basically due to harmonic
vibrations~\cite{footnote1}. For $T \gtrsim 0.55$, 
the approach of $s_{\rm c}$ to zero is milder and the bent is less sharp, 
indicating that the transition becomes a crossover.

We define the ideal glass transition point $c_K(T)$, or $T_K(c)$, as the
point at which $s_{\rm c}$ becomes zero.  As the temperature is lowered,
$c_K(T)$ decreases and in Fig.~4 we show the resulting phase diagram the
details of which will be discussed below. Finally we mention that the
presented results are for $N=300$. However, we have also simulated systems
with $N=150$ and found that the results do not depend significantly on $N$
(see SI for details).

\subsection{Overlap approach}
An alternative method to locate and characterize the thermodynamic
transition is to study the overlap $q_{\alpha\beta}$ between two
configurations $\alpha$ and $\beta$: $q_{\alpha\beta}=N^{-1} \sum_{i,j}
\theta(a-|\mathbf{r}_i^\alpha-\mathbf{r}_j^\beta|)$, where $\theta$
is the Heaviside function, $\mathbf{r}_i^\alpha$ is the position
of particle $i$ in configuration $\alpha$, and the length-scale $a$
is 0.3~\cite{Kob2013prl}.  RFOT predicts that at the glass transition
the average value $[\langle q \rangle]$ will increase quickly from
a small value in the fluid phase to a large value in the glass
phase~\cite{Cammarota2013jcp}.  Here $\langle \cdots \rangle$ and
$[\cdots]$ stand for the thermal and disorder averages, respectively.
We have computed the overlap distribution $P(q)$ using replica
exchange molecular dynamics~\cite{Kob2013prl} (see SI) and in Fig.~2 we
present $[P(q)]$ for $T=$0.7 and $0.45$. For $T=0.7$, $[P(q)]$ remains
single-peaked for all $c$, and the peak position shifts continuously
towards larger $q$ as $c$ increases, see Fig.~2(a). A qualitatively
different behavior is observed at $T=0.45$ (Fig.~2(b)): $[P(q)]$ is
single-peaked at low and high $c$, but has a double peak structure at
intermediate $c$, thus signalling the coexistence of the fluid and glass
phase, which indicates that the transition from the fluid phase to the
glass phase is first-order-like~\cite{Kob2013prl}.

The $c$-dependence of the average overlap $[\langle q \rangle]$ is shown
in Fig.~2(c). For high temperatures $T \gtrsim 0.6$, $[\langle q \rangle]$
smoothly increases with $c$, reflecting the continuous shift of the single
peak of $[P(q)]$ as shown in Fig.~2(a). For $T\lesssim 0.5$, $[\langle
q \rangle]$ shows a quick increase at intermediate values of $c$, in
agreement with the presence of the double peak structure seen in $[P(q)]$
at these $T$.  It suggests that a first-order-like transition, rounded by
finite size effect, takes place, in qualitative agreement with the result
for a system of harmonic spheres~\cite{Kob2013prl}. Note that, within the
accuracy of our data, we see that the amplitude of the (smeared out) jump
in $[\langle q \rangle]$ seems to vanish around $0.55 \lesssim T
\lesssim 0.6$, thus indicating that at around that temperature the line of
first order transition ends in a critical point that is second-order-like.

From this approach with the overlap, we can define the ideal glass
transition temperature $T_{K}^{(q)}(c)$ as the temperature at which the
skewness of $[P(q)]$ vanishes (see SI), and in Fig.~4 we have included the
resulting $T_{K}^{(q)}(c)$ as well. For small and intermediate $c$, we find
a very good agreement between $T_K(c)$ and $T_K^{(q)}$, thus showing that
the two very different approaches do give the same ideal glass transition
temperature. This is thus very strong evidence that at this temperature
the system does indeed undergo a thermodynamic phase transition
from the fluid to an ideal glass state. For temperatures above the
second-order-like critical point the curves $T_K(c)$ and $T_K^{(q)}(c)$
differ. We will rationalize this in the Discussion below. Finally we
mention that the curves $T_K(c)$ and $T_{K}^{(q)}(c)$ seem to extrapolate
in a smooth manner to the Kauzmann temperature of the bulk which has
been {\it estimated} from extrapolation~\cite{Sciortino1999prl}. This
shows that the present measurement of the bulk $T_K$ is compatible with
the one from previous estimates.

\begin{figure}[th]
\begin{center}
\hspace*{0.3cm}
\includegraphics[width=0.46\textwidth,clip]{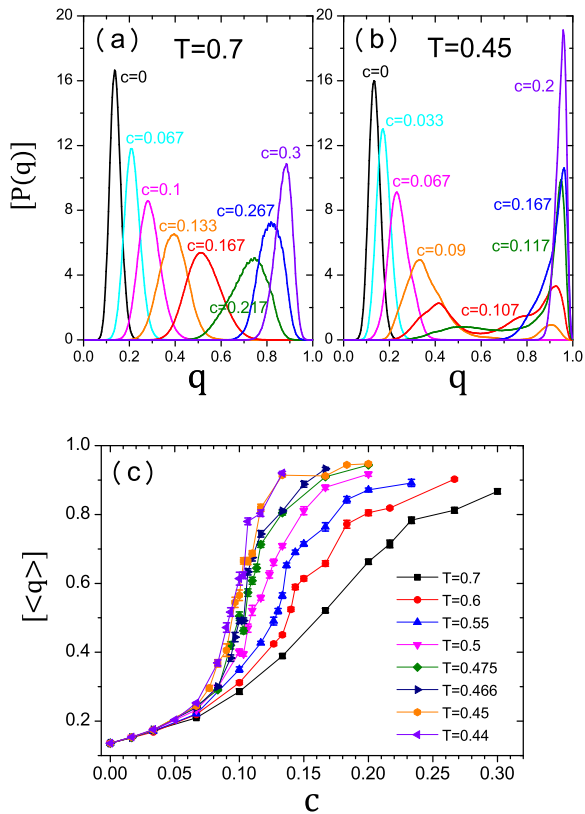}
\caption{
Distribution of the overlap $[P(q)]$ for $T=0.7$ (a) and $T=0.45$
(b). (c): The average overlap $[\langle q\rangle]$ obtained from $[P(q)]$
as a function of $c$ for several temperatures. The error bars have been
calculated using the jackknife method.
}
\end{center}
\label{fig-overlap}
\end{figure}
  
\subsection{Potential energy landscape and mode coupling temperature}

In the past, it has been found that the slow dynamics of glass-forming
systems is closely related to the features of the potential energy
landscape (PEL)~\cite{Sciortino2005d} and in the following we will use
these relations to characterize the relaxation dynamics of the pinned
system.

Figure~3(a) shows the $T-$dependence of the average inherent structure
energy  $[\langle \eIS\rangle]$.  For the bulk system, $c=0$, $[\langle
\eIS\rangle]$ is basically constant at high temperatures but then steeply
decreases below a crossover temperature $T\approx 1$, a temperature
which signals that the relaxation dynamics becomes strongly influenced
by the PEL~\cite{sastry1998,Brumer2004b}. As $c$ increases, the value
of $[\langle \eIS\rangle]$ at high $T$ moves steadily upward, which is
reasonable since the energy of the system is literally {\it pinned} at
the higher energy levels due to the presence of the pinned particles.
Concomitantly the crossover temperature increases with $c$ and the
crossover becomes smeared out, completely disappearing at the highest
$c$. The vanishing of this crossover with growing $c$ indicates thus
that the pinning affects qualitatively the nature of the PEL and of the
relaxation dynamics. For instance it is found that with increasing $c$
the fragility of the system decreases and shows at high $c$ an Arrhenius
dependence~\cite{Kob2013prl,Kim2009epl}.

In the inset of Figure~3(a), the low temperature behavior of $[\langle
\eIS\rangle]$ is shown.  It clearly demonstrates that $[\langle
\eIS\rangle]$ is inversely proportional to $T$ for all $c$.  According to
the energy landscape scenario, this is an indicator that the distribution
of $\eIS$ is Gaussian~\cite{Ruocco2004jcp}.

Another important quantity that connects the glassy dynamics
of a system with its PEL is the saddle index $K$, {\it i.e.},
the number of negative eigenvalues of the Hessian matrix at a
stationary point of the PEL. For bulk systems it has been found that
$K$ shows a linear dependence on $\eSP$, the bare energy of a saddle
point~\cite{angelani2000,Broderix2000}. Since the value of $\eSP$ at
which $K$ goes to zero (this value is often denoted as ``threshold
energy'' $\eTH$) corresponds to the average energy of the inherent
structures $\langle \eIS \rangle$ at the critical temperature of
mode-coupling theory, one can extract from the geometrical properties
of the PEL the value of $T_d$ without having to do any 
fit to dynamical data~\cite{angelani2000,Broderix2000}.

We use a standard method to determine numerically the energy and index
of saddles for the pinned system (see SI for details), and in Fig.~3(b)
we plot the average normalized saddle index $k=K/3N(1-c)$ as a function
of its corresponding $\eSP$. In agreement with previous studies of
the PEL, we find that $k$ decreases linearly as a function of $\eSP$
and hence we can obtain $\eTH(c)$ from a linear fit (included in the
figure as well).  We see immediately that $\eTH$ increases with $c$
and together with the data of Fig.~3(a) and $\eTH(c)= [\langle \eIS
\rangle](T_d(c))$, we can conclude that $T_d(c)$ increases with $c$.
The resulting $c-$dependence of $T_d$ is included in Fig.~4 as well and
will be discussed in the next section.

We have also evaluated $T_d$ by calculating the relaxation time
$\tau_\alpha$ from the time dependent density correlation function
(see {\it Material and Methods}) and by fitting $\tau_\alpha$ with
the MCT power-law $\tau_{\alpha} \simeq |T-T_{d}^{\fit}|^{-\gamma}$.
We find that the so obtained values of $T_{d}^{\fit}$ agree well
with those obtained from the PEL (see SI for details).  Note that
one needs several fit parameters to obtain $T_{d}^{\fit}$ from the
dynamic data, whereas basically no fit parameter are required for
$T_d$ from the PEL.  In  Fig.~4, the iso-$\tau_\alpha$ lines are also
plotted.  The graph shows that $\tau_\alpha$ quickly increases with
$c$ and that the lines asymptotically approach the $T_K$ line from
the high $T$-side.

\begin{figure}[tb]
\begin{center}
\includegraphics[width=0.48\textwidth,clip]{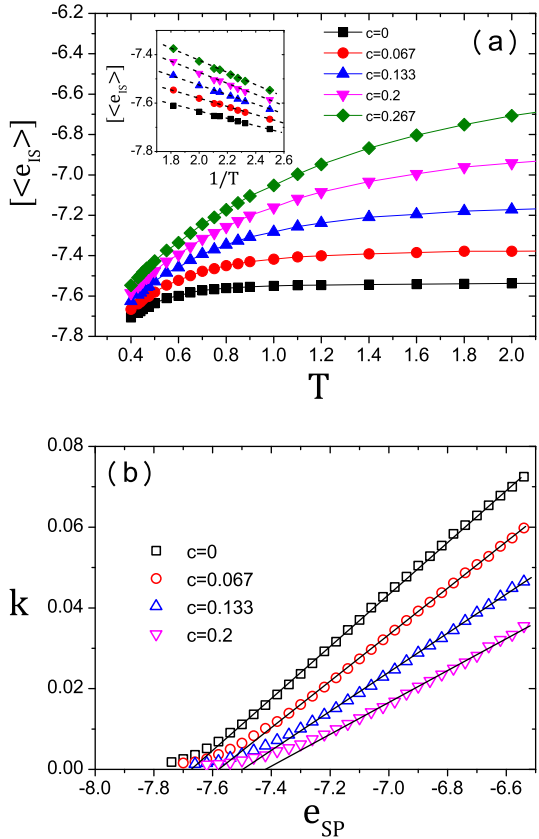}
\caption{
(a): $T$ dependence of the averaged inherent structure $[\langle \eIS
\rangle]$ for several values of $c$. Inset: Same quantity as a function
of $1/T$. (b): The average normalize saddle index, $k$, as a function of
its energy, $\eSP$, for several values of $c$. The threshold energy,
$\eTH$, is defined by a linear extrapolation from the high energy side.}
\end{center}
\label{fig-saddles}
\end{figure}  

\section{Discussion}
In Fig.~4 we summarize the results of the previous sections in the form
of a phase diagram in the $c$-$T$ plane. The ideal glass transition
lines $T_K(c)$ determined from $S_{\rm c}=0$ and $T_K^{(q)}(c)$ obtained
from $P(q)$ are plotted as filled circles and diamonds, respectively.
At low $T$ and $c$ the two temperatures basically coincide but around
$(T_{\rm c},c_{\rm c}) \approx (0.55, 0.16)$, they start to depart from
each other in that $T_K^{(q)}(c)$ increases continuously whereas $T_K(c)$
bends and its $c$-dependence becomes weaker. Note that this separation
occurs at the same point at which $[P(q)]$ changes from the bimodal to the
single-peaked shape (see Fig.~2), thus indicating that in this region
of parameter space the system has a second order like critical point.

The theoretical calculations for a mean-field spin glass model show that
$T_K(c)$ should terminate at a finite $c$ and that this end point is a
critical point of the universality class of the random field Ising 
model~\cite{Cammarota2012pnas,Cammarota2013jcp,Cammarota2013epl,Franz2013jsmte3}.
Furthermore the theory predicts that at this end point the coexistence
line $T_K(c)$  and the dynamical line $T_d(c)$ merge. Figure~4 shows
that this prediction is indeed compatible with our data in that the two
lines do cross near $(T_{\rm c},c_{\rm c})$.

From the figure we also recognize that, beyond the endpoint, $T_K(c)$
obtained from $S_{\rm c}=0$ almost matches with $T_d(c)$.  This result
is reasonable since in this range of $T$ and $c$ the particles are
strongly confined by the labyrinthine structure imposed by the pinned
particles, {\it i.e.},  the system resides mainly at the bottom of
a free energy minimum where both the configurational entropy and the
number of the saddles vanish simultaneously. In contrast to $T_K(c)$,
$T_K^{(q)}(c)$ raises continuously even beyond the endpoint. This can be
understood by recalling that this line is defined by the points at which
the skewness becomes zero, {\it i.e.},  the point at which $[P(q)]$
changes from being left-skewed to right-skewed. Since this change of
sign in the skewness occurs also in the region of the phase diagram
beyond the critical point, the line $T_K^{(q)}$ will extend into that
region, similar to the Widom line present in a standard liquid-gas
transition~\cite{Widom1965jcp,Xu2005pnas}.

As it is evident from Fig.~1(a), $T_K$ defined as the point at which
$S_{\rm c}=S-S_{\vib}=0$ becomes somewhat ill-defined beyond the endpoint
($T \gtrsim 0.55$ in Fig.~1(a)), since the crossover from the fluid
to the glass phase becomes broad. In order to estimate the effect of
this ambiguity, we have included in Fig.~4 also $T_K^{\prime}$ defined
by the linear extrapolation $S_{\rm c}$ of Fig.~1(b) from the {\it low} $c$ side
(open circles). We see that below the end point $T\lesssim 0.5$,
$T_K^{\prime}$ is indistinguishable from $T_K$ whereas beyond the
end point $T \gtrsim 0.6$ they bifurcate and $T_K^{\prime}$ becomes
comparable with $T_K^{(q)}$.

\begin{figure}[thb]
\begin{center}
\includegraphics[width=0.48\textwidth,clip]{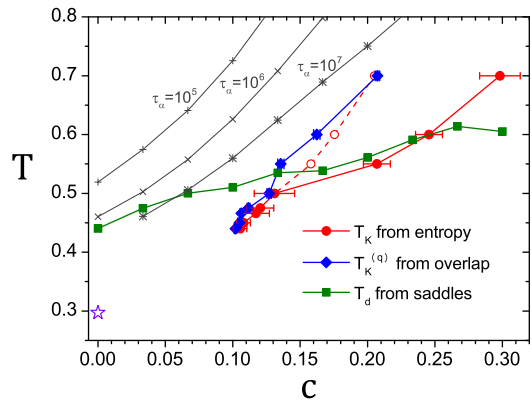}
\caption{
The phase diagram of the randomly pinned system.  The filled circles show
the ideal glass line $T_K$ at which the configurational entropy vanishes.
The diamonds show $T_K^{(q)}(c)$ determined from the skewness of $[P(q)]$.
The squares are dynamic transition points $T_d(c)$.  The open circles show
the ideal glass line $T_K^{\prime}$ determined by the linear extrapolation
of $S_{\rm c}(c)$ to vanish from Fig.~1(b).  The iso-relaxation-times
are drawn by $+$, $\times$, and $\ast$.  The star denoted at $c=0$ is
a putative ideal glass transition point $T_K \approx 0.3$ for the bulk
reported in Ref.~\cite{Sciortino1999prl}.
}
\end{center}
\label{fig-phasediagram}
\end{figure}  

To the best of our knowledge, the present study is the first report
of a system in finite dimensions that shows the existence of an
ideal glass state {\it in equilibrium}, {\it i.e.}, a state in
which the configurational entropy is zero at a finite $T$. The
Kauzmann temperatures reported in the past have all relied on
somewhat questionable extrapolation procedures, leaving thus
room for debate over the very existence of a thermodynamic
transition~\cite{Tarjus2010,Chandler2010arpc,Garrahan2014pre}.

Our findings are inconsistent with recent simulation studies in
which the $T-$and $c-$dependence of the relaxation dynamics has been
studied~\cite{Chakrabarty2014}.  In Ref.~\cite{Chakrabarty2014},
the structural relaxation time $\tau_\alpha$ was fitted with the
Vogel-Fulcher relation $\tau_\alpha \sim \exp[A/(T-T_0)]$, with $T_0$
as fit parameter. This relation can be directly derived from the
Adam-Gibbs as well as the RFOT theory, both of which assert that
$\ln\tau_\alpha \propto 1/TS_{\rm c}$, and thus the Vogel-Fulcher
temperature $T_0$ is predicted to be identical to $T_K$. Furthermore
the authors of Ref.~\cite{Chakrabarty2014} fitted their data also with
the MCT power-law $\tau_{\alpha} \simeq |T-T_{d}|^{-\gamma}$ in order
to determine the $c-$dependence of $T_d$. It was found that while
$T_{d}$ increases moderately with $c$, $T_0$ remains constant. We
have plotted the relaxation time $\tau_\alpha$ as a function of
$S_{\rm c}$ for finite $c$ and found that the Adam-Gibbs relation is
violated (see SI for details). Thus we conclude that in the case
of pinned systems one cannot deduce the Vogel-Fulcher law from the
Adam-Gibbs relation.

Since the results presented here are all obtained in thermodynamic
equilibrium without referring to any kind of extrapolation, we are
confident that the phase diagram presented in Fig.~4 does indeed reflect
the properties of the system and is not an artifact of the analysis. 
A further evidence that the simulated system is really in equilibrium is 
the observation that the entropy obtained by thermodynamic
integration from the high temperature limit matches with that
obtained from the low temperature side (via harmonic approximation) in the glass phase. 
It is also reassuring that all three methods, the thermodynamic integration
(vanishing entropy), the overlap distribution (the discontinuous jump
of $q$), and the geometric change of the PEL, consistently point to
the same end point, thus giving strong evidence that this point really
exists. Also suggestive is that each combination of pairs amongst three
methods are compatible beyond the end point, which is reminiscent of
the Widom line in the standard gas-liquid phase transition.

At this stage we can conclude that the phase diagram as predicted by
the RFOT theory is confirmed at least qualitatively. What remains
to be done is to probe the relaxation dynamics in the vicinity of
the critical end point since one can expect that this dynamics is
rather unusual~\cite{Nandi2014prl} and to establish its universality
class~\cite{Franz2013jsmte3,Biroli2014prl}. Furthermore, it will also be
important to see whether the predicted phase diagram can also be observed
in real experiments. Although this will be not easy, for certain systems
such as colloids or granular media it should be possible.

\section{Materials and Methods}
\subsection{Model} 

The system we use is a binary mixture of Lennard-Jones
particles~\cite{Kob1995c}.  Both species $A$ and $B$ have the same mass
and the composition ratio is $N_A : N_B = 80 : 20$.  The interaction
potential between two particles is given by $v_{\alpha \beta}(r)
= 4\epsilon_{\alpha \beta}\{ (r/\sigma_{\alpha \beta} )^{12} -
(r/\sigma_{\alpha \beta})^{6}\}$, where $\alpha, \beta \in \{A,
B\}$.  We set $\epsilon_{AA}=1.0,\epsilon_{AB}=1.5,\epsilon_{BB}=0.5,
\sigma_{AA}=1.0,  \sigma_{AB}=0.8$ and $\sigma_{BB}=0.88$.  $v_{\alpha
\beta}(r)$ is truncated and shifted at $r=2.5\sigma_{\alpha \beta}$.
We show energy in units of $\epsilon_{AA}$, with the Boltzmann constant
$k_B=1$, and length in units of $\sigma_{AA}$. Time units are defined by
Monte Carlo sweeps (see below).  Simulations are performed at constant
density $\rho\approx 1.2$.  The number of particles is $N=150$ and 300,
and most of the results in the present study are for $N=300$.

\subsection{Making pinned configurations}

The configuration of the pinned particles is generated by making
first a replica exchange run for the bulk system, {\it i.e.},  $c=0$,
using 8 replicas~\cite{Hukushima1996jpsj}. This allows us to generate
relatively quickly many equilibrium configurations that are completely
independent, {\it i.e.}, between consecutive configurations the
mean squared displacement of a tagged particle is more than 100.
Next we use a ``template'' to identify the $cN$ particles that will be
permanently pinned. Details on how to create the template can be found
in Ref.~\cite{Kob2013prl}. This approach has the advantage that the
pinned particles cover the space in a relatively uniform manner, thus
avoiding the creation of dense regions or large empty regions and hence
suppressing strong sample-to-sample fluctuations of the thermodynamic
properties of the system.

\subsection{Simulation methods}
- Thermodynamics:
In order to sample thermodynamic properties efficiently at
low $T$ and large $c$ region, we use the replica exchange
method~\cite{Hukushima1996jpsj}.  The maximum number of replicas
is 24.  More detail is presented in the supplemental information and in
Ref. \cite{Kob2013prl}.  The total CPU time to obtain the presented results is 
about 580 years of single core time.
\\
- Dynamics:
We use the Monte Carlo (MC) dynamics simulation to calculate dynamical
observables~\cite{Berthier2007i}. The rule of the MC dynamics is the
following: In an elementary move, one of the $(1-c)N$ unpinned particle
is chosen at random. Then the particle is displaced at random within a
cubic box of linear size $\delta= 0.15$  and the standard Metropolis
rule is used to decide whether or not the move is accepted. One MC
step consists of $(1-c)N$ such attempts and we set this as a unit of
time scale.  The relaxation time $\tau_{\alpha}$ is determined by $F^A(k,
\tau_{\alpha})= e^{-1}$, where $F_s^A(k, t)$ is the self part of the
intermediate scattering function of the free particles of species $A$ for
the wave-vector $k$ at the peak of the corresponding structure factor. We
have averaged over 30 different realizations of pinned particles to
calculate $F^A(k, t)$.

\subsection{Entropy}
In order to calculate the entropy $S(c,T)$ of the pinned system, we have
first determined the entropy of the system for a given configuration
of pinned particles, and then taken the average over the realization of
the configuration of pinned particles. For this we have calculated the
potential energy at temperatures ranging from the target temperature
up to the ideal gas limit at $T=\infty$, while keeping the pinning
configuration fixed.  We have evaluated the entropy of the system with
that pinning configuration using thermodynamic integration. For this
integration, we have used a grid in the inverse temperature $\beta = 1/T$
of width $\Delta \beta$ that ranged between 0.01 to 0.1, depending on the
temperature, and integrated the potential energy as a function of $\beta$.
Special care was taken in the very high temperature regime, in order to
accurately and rapidly achieve the convergence to high temperature ideal
gas limit. The high temperature expansion of the potential energy of
Lennard-Jones fluid can be written as 
$U = A\beta^{-1/4} + B\beta^{-2/4}
+ C\beta^{-3/4}$~\cite{Coluzzi2000b}.  We have used simulations at very
high temperatures to determine the coefficients $A$, $B$, and $C$,
and 
carried out then the thermodynamic integration analytically.

\subsection{Analysis of the saddles}
To locate the saddles of the potential energy landscape of the system, we
have made a minimization of the squared gradient potential $W=\frac{1}{2}
|\nabla U |^2$\cite{angelani2000,Broderix2000}. Minimization of $W$ is
performed by the BFGS method~\cite{NW1999}. Similar to the minimization of
$U$ used to calculate the inherent structures, $W$ includes the position
of the pinned as well as unpinned particles, but only the position of
the latter are optimized.  After having located a saddle with energy
$\eSP$, the Hessian matrix was diagonalized and we counted the fraction
of negative eigenvalues $k(\eSP)$.  The raw data is shown in the
SI and in Fig.~3 (b) we present the average index as a function of $\eSP$.
The threshold energy $\eTH$ is defined by $k(\eTH)=0$.

\begin{acknowledgments}
We thank G. Biroli, C. Cammarota, D. Coslovich, and K. Kim for helpful discussions.
MO acknowledge the financial support by Grant-in-Aid for JSPS Fellows (26.1878).
WK acknowledges the Institut Universitaire de France.
AI acknowledges JSPS KAKENHI No. 26887021. 
KM and MO acknowledge KAKENHI 
No. 24340098, 
25103005, 
25000002, 
and the JSPS Core-to-Core Program.
The simulations have been done in Research Center
for Computational Science, Okazaki, Japan, at the HPC@LR, and the CINES (grant c2014097308).
\end{acknowledgments}

\begin{center}
{\bf \large Supplementary Information}
\end{center}
\newcommand{\bvec}[1]{\mbox{\boldmath $#1$}}
\newcommand{\ave}[1]{\langle #1 \rangle}
\setcounter{section}{0}

\section{Replica exchange method}

\begin{figure}[htbp]
\begin{center}
\includegraphics[width=0.9\columnwidth]{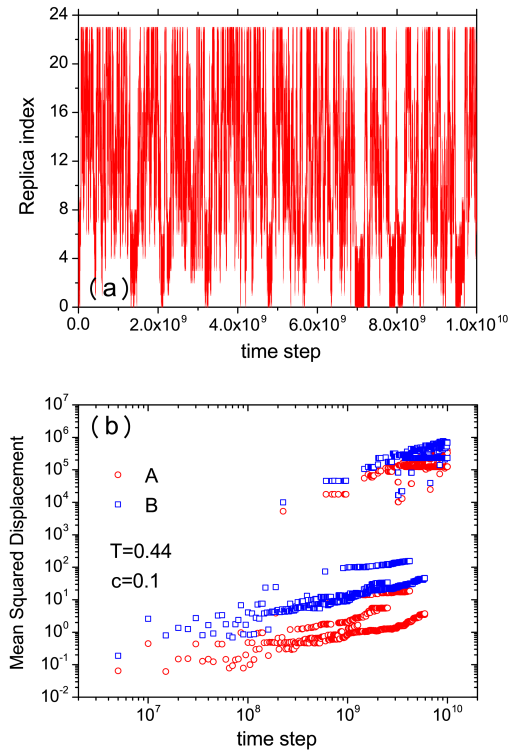}
\caption{
(a): Time dependence of the trajectory of a typical replica in temperature space for $T=0.44$, $c=0.1$.
(b): The mean squared displacement of $A$ and $B$ particles for $T=0.44$, $c=0.1$.
}
\label{fig:1}
\end{center}
\end{figure}

Here we describe the details of the replica exchange method
which we have employed in this study, which is basically the
repetition of the description in Ref.~\cite{SI_Kob2013prl} and in
its supplemental information.  We have used the parallel tempering
algorithm~\cite{SI_hukushima1996exchange,SI_yamamoto2000replica} using
up to 24 replicas.  In this approach, one simulates simultaneously
several copies of the system {\it i.e.}, the same Hamiltonian but each
replica is at a different temperature. Using a Boltzmann criterion and
the detailed balance condition, we periodically attempt to interchange
the configurations of two replicas at different temperatures.  Hence each
replica makes a random walk in the temperature space.  Due to the fast
relaxation at high $T$, this will lead to an efficient relaxation of the
system also at low $T$.  The smallest difference in temperature between
the two neighboring replicas was $\Delta T =0.009$, which guarantees a
good overlap of the two potential energy distributions.  Attempts to
switch the two neighboring replicas have been made every 50000 time
steps. We have checked that each parallel tempering run has indeed reached
equilibrium by following any given replica and making sure that all
temperatures have been sampled sufficiently.  A typical path of a replica
in temperature space is shown in Fig.~SI-\ref{fig:1}(a).  The parallel
tempering algorithm indeed allows for the system to be equilibrated
down to $T = 0.44$ even when the concentration of pinned particles,
$c$,  is large.  This can be recognized from Fig.~SI-\ref{fig:1}(b),
where we show the mean squared displacement (MSD) of the particles
(distinguishing the type $A$ and $B$ particles) as a function of time.
The figure shows that at sufficiently long times the MSD becomes very
large, indicating that the particles do indeed move through the box
also at high values of $c$, even if their relative arrangement does not
change much, {\it i.e.}, one is in the glass state. To calculate physical
quantities, we have followed a given replica in the temperature space
and considered only the time intervals at which this replica was at the
target temperature.

\section{Entropy of the pinned systems}

For our numerical calculation of the entropy, we employed
the thermodynamic integration technique.  Although this
technique is well-known and has long been applied to bulk
systems~\cite{SI_sciortino1999inherent,SI_sastry2000evaluation,SI_coluzzi2000lennard-jones},
its extension to pinned systems is not trivial.  In this section,
we summarize the ideas and present the details of our calculation.
We follow the notations introduced in the Method and Materials section.

\subsection{Definitions}

Our starting system is a binary mixture of type A and B particles.
Since we consider a random pinning of this system, our system is now
composed of four different types of particles: $cN_A$ pinned particles
of type A, $cN_B$ pinned particles of type B, $M_A = (1-c)N_A$ unpinned
particles of type A, and $M_B = (1-c)N_B$ unpinned particles of type B.
The total pinned and unpinned particle numbers are $cN = cN_A + cN_B$
and $M = M_A + M_B$, respectively.  We denote the coordinate of
pinned particles by $\bvec{S} = (\bvec{s}_1, \cdots, \bvec{s}_{cN})$
and unpinned particles by $\bvec{R} = (\bvec{r}_1, \cdots, \bvec{r}_{M})$.

First we consider the thermodynamics of the system with a given configuration
of pinned particles $\bvec{S}$.  In this situation, the partition
function is defined as

\begin{eqnarray}
\tilde{Z}(\bvec{S},\beta) = \frac{1}{\Lambda^{3M} M_A! M_B!} \int d \bvec{R} \ \exp[-\beta U]. 
\end{eqnarray}

\noindent
and the thermal average of any variable is 

\begin{eqnarray}
\ave{A}_{\bvec{S},\beta} \equiv \frac{1}{\tilde{Z}(\bvec{S},\beta) \Lambda^{3M} M_A! M_B!}
\int d \bvec{R} \ A \exp[-\beta U],  
\end{eqnarray}

\noindent
where $\Lambda$ is the thermal de Broglie wavelength, $U$ the potential
energy, and $\beta=1/k_BT$ characterizes the temperature of the system.
The free energy and the entropy are now written as

\begin{eqnarray}
\tilde{F}(\bvec{S},\beta) & = & - \frac{1}{\beta} \log \tilde{Z}(\bvec{S},\beta), \\ \
\tilde{S}(\bvec{S},\beta) & = &\beta \ave{H}_{\bvec{S},\beta} - \beta
\tilde{F}(\bvec{S},\beta),
\end{eqnarray}

\noindent
where $H$ is the Hamiltonian of the unpinned particles.

In this work, we generate pinned particles configurations by fixing a
fraction $c$ particles in the equilibrium configurations, and study
the thermodynamics after the disorder average over the realizations
of pinned particles.  When the pinned particles are generated at a
temperature $\beta'$, their distribution function is given by

\begin{equation}
P_{\rm pin}(\bvec{S}, \beta') = 
\frac{1}{Z(\beta') \Lambda^{3N} N_A! N_B!} \int d \bvec{R} \ \exp[-\beta' U],  
\end{equation}

\noindent
where $Z(\beta')$ is the partition function of the bulk system at the
temperature $\beta'$. Accordingly the thermodynamic quantities to be calculated 
are given by

\begin{eqnarray}
[\ave{A}_{\bvec{S},\beta}]_{\beta'} \equiv \int d\bvec{S} \ P_{\rm pin}(\bvec{S}, \beta') \ave{A}_{\bvec{S},\beta}.
\end{eqnarray}

\noindent
The subscript $\beta$ of the thermal average $\ave{\dots}$ indicates that
the thermal average for unpinned particles are taken at $\beta$, while
the subscript $\beta'$ of the disorder average $[\dots]$ indicates that
the pinned particles configurations are obtained from the equilibrium
configurations at the temperature $\beta'$.  Because we are interested
mainly in the case $\beta = \beta'$, the free energy and the entropy of
the system is now written as

\begin{eqnarray}
F(c,\beta) = [\tilde{F}(\bvec{S},\beta)]_{\beta}, \ \ \ 
S(c,\beta) = [\tilde{S}(\bvec{S},\beta)]_{\beta}. 
\end{eqnarray}

\noindent
Note that when $\beta = \beta'$, the thermodynamic average of a mechanical
variable $A$ becomes exactly

\begin{eqnarray}
[\ave{A}_{\bvec{S},\beta}]_{\beta} = \frac{1}{Z(\beta) \Lambda^{3N} N_A! N_B!} 
\int d\bvec{S} d \bvec{R} \ A \exp[-\beta U],  \label{equiv}
\end{eqnarray}

\noindent
which is nothing but the thermodynamic average for the bulk system at
the temperature $\beta$.  This equivalence is the well-known fact for
this type of the pinned system~\cite{SI_Scheidler2004jpcb}.

\subsection{The thermodynamic integration}

\subsubsection*{Formulation}

To calculate the entropy $S(c,\beta)$ from simulations, we employ
the thermodynamic integration (TI) method.  The TI method that is
most frequently applied to Lennard-Jones particles is to connect
the ideal gas state and the state of interest by a combination of a
compression path (increasing density) and a cooling path (decreasing
temperature)~\cite{SI_sciortino1999inherent,SI_sastry2000evaluation}.
However for pinned fluids, it is not clear how compression should
be defined.  Therefore we use only the cooling path from the ideal
gas limit $T=\infty$ to the target temperature without changing the
density.  This version of the TI has been used for bulk Lennard-Jones
particles~\cite{SI_coluzzi2000lennard-jones}, and here we apply it to the
pinned system.

We first apply the TI to the entropy of a given pinned particles
configuration, and then take the disorder average over realizations.
The entropy of the system with the pinned particles $\bvec{S}$ at the
target temperature $\beta^{\ast}$ can be expressed as

\begin{eqnarray}
\tilde{S}(\bvec{S}, \beta^{\ast}) = \tilde{S}(\bvec{S}, 0) + 
\beta^{\ast} \ave{U}_{\bvec{S},\beta^{\ast}} - \int^{\beta^{\ast}}_0 d\beta \ \ave{U}_{\bvec{S},\beta}. \label{ti0}
\end{eqnarray}

\noindent
$\tilde{S}(\bvec{S}, 0)$ is the entropy of $M$ unpinned particles at
$\beta = 0$.  Since the interactions between the particles including
pinned and unpinned particles become irrelevant if $\beta = 0$, this term is
nothing else than the ideal gas entropy of bulk $M$ particles.  The thermal
average of the potential energy can be decomposed as

\begin{eqnarray}
\ave{U}_{\bvec{S},\beta} = \ave{U_{\rm up}}_{\bvec{S},\beta} + U_{\rm p}(\bvec{S}) 
+ \ave{U_{\rm int}}_{\bvec{S},\beta},
\end{eqnarray}

\noindent
where $U_{\rm up}$, $U_{\rm p}$, and $U_{\rm int}$ 
are the potential energies of unpinned particles, pinned particles and the interaction
between pinned and unpinned particles, respectively.  Here we use
the fact that $U_{\rm p}$ is solely dependent on the configuration of
pinned particles and free from the thermal average of unpinned particles.
Plugging these into Eq.~(\ref{ti0}) and taking the disorder average over
realizations, we get the final expression

\begin{equation}
\begin{aligned}
S(c,\beta^{\ast}) 
&=
 M\left(1 - \log \frac{M}{V}\right) - M_A \log \frac{M_A}{M} 
\\
& - M_B \log \frac{M_B}{M} - 3M \log \Lambda + \frac{3M}{2} 
\\ 
&
+ \beta^{\ast} [\ave{U_{\rm up}}_{\bvec{S},\beta^{\ast}} + \ave{U_{\rm int}}_{\bvec{S},\beta^{\ast}}]_{\beta^{\ast}} 
\\ 
&
- \int^{\beta^{\ast}}_{0} d\beta \ [\ave{U_{\rm up}}_{\bvec{S},\beta} + 
\ave{U_{\rm int}}_{\bvec{S},\beta}]_{\beta^{\ast}}. 
\end{aligned}
\label{eq:S_final}
\end{equation}

\noindent
In the integral, the temperature for the thermal average runs from
$\beta=0$ to $\beta=\beta^{\ast}$, while the temperature for the disorder
average over realizations of the pinned particles is fixed to be the
target temperature $\beta^{\ast}$.  Thus we cannot use the relation
Eq.~(\ref{equiv}) to evaluate the integrands. Instead we need to
directly calculate the potential energy at $\beta \in [0,\beta^{\ast}]$
under a given pinned particles configuration $\bvec{S}$ and take the
disorder average over realizations.

\subsubsection*{Implementation}

\begin{figure}[htbp]
\begin{center}
\includegraphics[width=0.9\columnwidth]{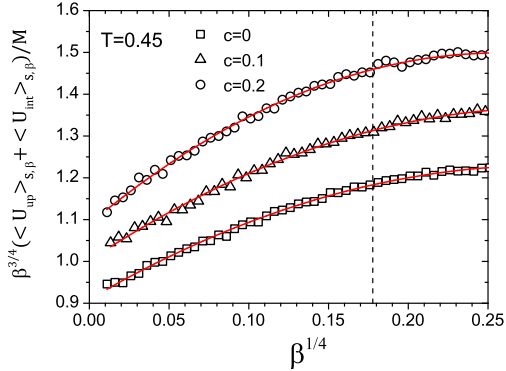}
\caption{
$\beta^{3/4}( \ave{U_{\rm up}}_{\bvec{S},\beta} + \ave{U_{\rm
int}}_{\bvec{S},\beta} )/M$ as a function of $\beta^{1/4}$ for very high
$T$ ($>250$). The raw data (symbols) are fitted well by the fitting curves
(red solid lines). The vertical dashed line indicates the temperature
above which we use the analytical integration. Note that we show here only
the data for one realization of the pinned particles since the curves will
depend on the realization.
}
\label{fig:2}
\end{center}
\end{figure}

The integral in Eq.~(\ref{eq:S_final}) is decomposed into three
temperature regimes and each of them is evaluated separately.  
(1)
For low temperature regime, we calculate the potential energy from the
configurations generated by the parallel tempering calculations outlined
in Sec. I.  From each trajectory, we obtain the thermal average of the
potential energies $U_{\rm up}$ and $U_{\rm int}$ at different temperatures
with a given pinned particles configuration.  We employ the Simpson's rule
to evaluate the integral and take the disorder average over realizations.
(2) Above the highest temperature in the parallel tempering calculations,
we use standard Monte Carlo method to calculate the potential energy.
We slice the temperature into the grids with the width $\Delta \beta
= 0.001 - 0.01$ depending on the temperature range and calculate the
thermal average of $U_{\rm up}$ and $U_{\rm int}$ at each temperature
for a given pinning configuration. We employ the Simpson's rule to
evaluate the integral and take the disorder average over realizations.
(3) We take a special care about the integration at very high temperature
$\beta < 0.001$, since the potential energies $U_{\rm up}$ and $U_{\rm
int}$ diverge in the high temperature limit.  To accurately calculate
the integral, we first fit the potential energy data to a polynomial
function then analytically integrate the function.  Considerations on
the high temperature expansion shows that the potential energy of the
Lennard-Jones particles at high temperature behaves as

\begin{equation}\label{eq:U_expansion}
\ave{U} = A \beta^{-1/4} + B \beta^{-1/2} + C \beta^{-3/4} + \mathcal{O}(1),
\end{equation}

\noindent
where $A$, $B$ and $C$ are constants~\cite{SI_coluzzi2000lennard-jones}.
We thus fit the data of $\ave{U_{\rm up}}_{\bvec{S},\beta} +
\ave{U_{\rm int}}_{\bvec{S},\beta}$ with expression~(\ref{eq:U_expansion})
to obtain these constants and integrate it analytically.  In
Fig.~SI-\ref{fig:2}, we show the raw data of $\beta^{3/4}(
\ave{U_{\rm up}}_{\bvec{S},\beta} + \ave{U_{\rm int}}_{\bvec{S},\beta}
)/M$ and the fitting function Eq.~(\ref{eq:U_expansion}) as a function
of $\beta^{1/4}$, 
confirming that Eq. (\ref{eq:U_expansion}) holds well.

\subsection{Vibrational entropy}

We next summarize the method to calculate the harmonic vibrational
entropy of the pinned system. We consider the system that weakly vibrates
around an inherent structure (IS). If we denote by $\delta \bvec{r}_i$
the displacement of the $i$-th particle from its position in the inherent
structure, the potential energy can be approximated well as

\begin{equation}
U \approx U_{\rm IS}(\bvec{S}) 
+ \frac{1}{2} \sum_{i, j}^M \frac{\partial^2 U}{\partial \bvec{r}_i \partial \bvec{r}_j} \delta \bvec{r}_i \delta \bvec{r}_j.
\end{equation}

\noindent
It is important to realize that only the derivative of the potential
energy respect to the coordinates of unpinned particles should be taken into
account, not including the ones of pinned particles (But of course $U$
will depend on the positions of the pinned and unpinned particles). 
Thus the size of the Hessian matrix is $(3M \times 3M)$. Introducing the eigenvalues 
$\lambda_1, \cdots, \lambda_{3M}$ of the Hessian, the harmonic vibrational
entropy of the given inherent structure with a given pinned particle
configuration can be written as

\begin{eqnarray}
\tilde{S}_{\rm IS, vib}(\bvec{S},\beta) = \sum_{a=1}^{3M} \left\{ 1- \log (\beta \hbar \sqrt{\lambda_a/m}) \right\}.
\end{eqnarray}

\noindent
Note that the eigenvalues $\lambda_a$ depends on the choice of the
inherent structure and the pinned particle configuration.  Taking the
average of $\tilde{S}_{\rm IS, vib}(\bvec{S},\beta)$ over realizations
of pinned particle configurations and the inherent structures, we finally 
obtain the harmonic vibrational entropy of pinned system $S_{\rm vib}(c,\beta)$.

In practice, we have sampled the inherent structure by minimizing
the potential energy of instantaneous configurations obtained in the
simulations.  For this calculation we used the conjugate-gradient method.
Note that the pinned particles were frozen during the minimization
process and only the coordinates of the unpinned particles are optimized.

\subsection{Finite size effect}

\begin{figure}[htbp]
\begin{center}
\includegraphics[width=0.9\columnwidth]{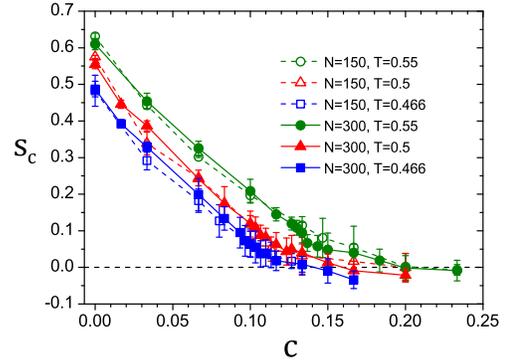}
\caption{
Finite size effect of the configurational entropy for $N=150$ and $N=300$.
}
\label{fig:3}
\end{center}
\end{figure}

In Fig.~SI-\ref{fig:3}, we show $S_{\rm c}$ for $N=150$ and 300 for
several state points.  We find that finite size effects are small,
at least for these two system sizes.

\section{Obtaining the Kauzmann temperature from the overlap}

We have determined the averaged distribution function $[P(q)]$ of the
overlap $q$ as explained in the main text. Subsequently we have calculated the
skewness $\gamma(c,T)$ as

\begin{equation}
\gamma = \cfrac{ \int_0^1 dq [P(q)](q-[\langle q\rangle ])^3}{ \left( \int_0^1 dq [P(q)] (q- [\langle q\rangle])^2\right)^{3/2} } \quad .
\end{equation}

\begin{figure}[htbp]
\begin{center}
\includegraphics[width=0.9\columnwidth]{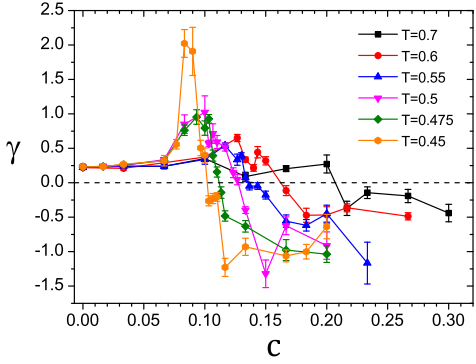}
\caption{
The skewness $\gamma$ of the distribution function $[P(q)]$.
The horizontal dashed line is the zero-axis and used to define $T_K^{(q)}$.
The error bars have been calculated using the jackknife method.
}
\label{fig:4}
\end{center}
\end{figure}

The $T-$and $c-$dependence of $\gamma$ is shown in Fig.~SI-\ref{fig:4}. The
point at which $\gamma$ is zero is used to define the Kauzmann temperature
$T_K^{(q)}$ shown in Fig.~4 of the main text.

\vspace*{10mm}

\section{Saddle points and dynamic transition point}

\begin{figure}[ht]
\begin{center}
\includegraphics[width=0.9\columnwidth]{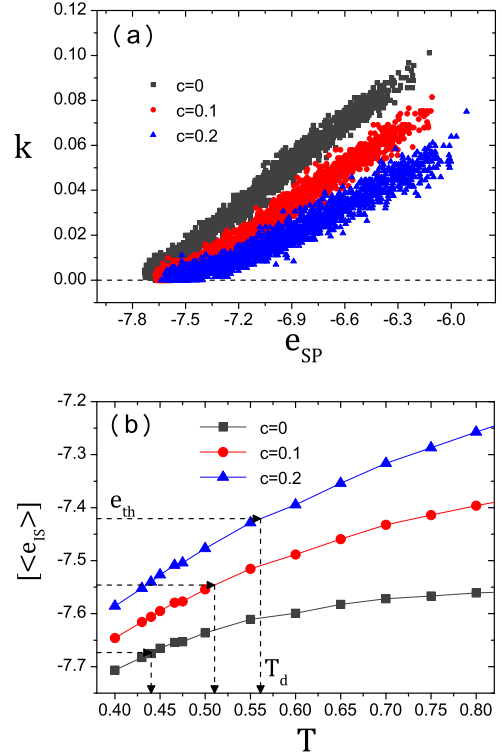}
\caption{
(a): The scatter plot of $k$ vs $\eSP$.
(b): The $[\langle \eIS \rangle]$ as a function of $T$.
The horizontal and vertical arrows indicate the location of $\eTH(c)$ and $T_d(c)$, respectively.
This plot is used to map $\eTH(c)$ to $T_d(c)$.
}
\label{fig:5}
\end{center}
\end{figure}  

In Fig.~SI-\ref{fig:5}(a), we show the original data of Fig.~3(b) of the
main text.  This scatter plot shows the normalized saddles index $k$
as a function of the energy at the saddle, $\eSP$, obtained from the
minimization of the square gradient potential $W=\frac{1}{2} |\nabla U
|^2$~\cite{SI_ADRSS2000,SI_BBCZG2000}.  Fig.~3(b) in the main text has
been obtained by averaging over this scatter data for a given $\eSP$.
The threshold energy $\eTH$ is extracted from Fig.~3(b) of the main text
as a point at which the averaged $k$ vanishes.  Mapping of  $\eTH(c)$
to the dynamic transition temperature $T_d(c)$ can be done by plotting
the temperature dependence of the inherent structures $[\langle \eIS
\rangle]$.  The reason why we use the inherent structures $[\langle
\eIS \rangle]$ instead of $[\langle \eSP \rangle ]$ is that in practice
one can evaluate the $T-$dependence of $[\langle \eIS \rangle]$ with
higher precision than that of $[\langle \eSP \rangle ]$.  Note that
using $[\langle \eSP \rangle]$ would in fact give the same result since
it is expected to be very close to $[\langle \eIS \rangle ]$ in the low
temperature regime.  In Fig.~SI-\ref{fig:5}(b) we show the $T$-dependence
of $[\langle \eIS \rangle]$ for several $c$'s. The inherent structures
are found to be a monotonic function of $T$ for all $c$'s and therefore
$\eTH$ can be uniquely mapped to $T_d$.

\begin{figure}[b]
\begin{center}
\includegraphics[width=0.9\columnwidth]{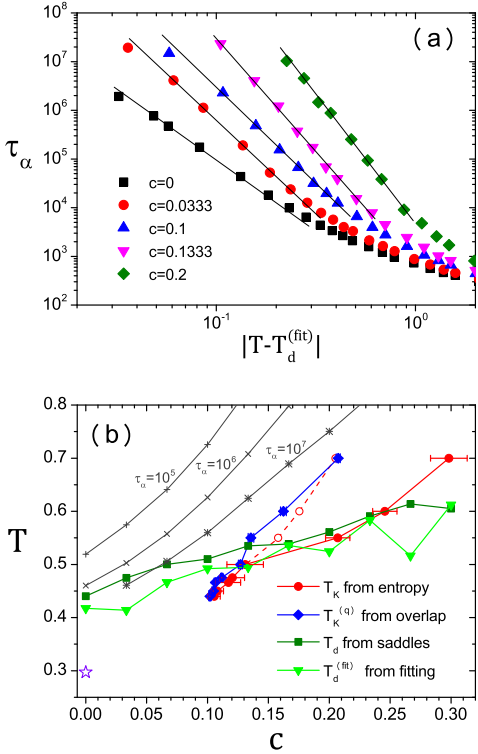}
\caption{
(a): Test of the validity of the MCT power-law fit to $\tau_{\alpha}$
where $T_d^{\rm (fit)}$ is a fitting parameter.
The solid lines indicate the MCT power-law.
(b): Phase diagram including the $T_d^{\rm (fit)}$(c) line.
}
\label{fig:S1}
\end{center}
\end{figure}

\section{Dynamic transition temperatures}

The dynamic transition temperatures $T_d$ can be
evaluated by two methods: the dynamic route and the static
one~\cite{SI_ADRSS2000,SI_BBCZG2000,SI_Gotze2009}.  In the dynamic route,
$T_d$ is obtained directly by fitting the relaxation time $\tau_\alpha$
from the time-dependent correlation functions, such as the scattering
function $F_s^A(k,t)$, to the MCT power-law $\tau_{\alpha}\propto |T-T^{\rm
(fit)}_d|^{-\gamma}$. Thus in this case one has $T^{\rm (fit)}_d$,
$\gamma$, and the prefactor are fit parameters, all of which will depend
on $c$. In the present paper, we have adopted the alternative static route
to obtain $T_d$ defined as a point at which the saddles of the potential
energy landscape vanish.  For bulk systems, $c=0$, it has been found that
$T^{\rm (fit)}_d$ from fitting (dynamic) agrees very well with $T_{d}$
as determined from the saddles (static) \cite{SI_ADRSS2000,SI_BBCZG2000}.

\begin{figure}[tb]
\begin{center}
\includegraphics[width=0.9\columnwidth]{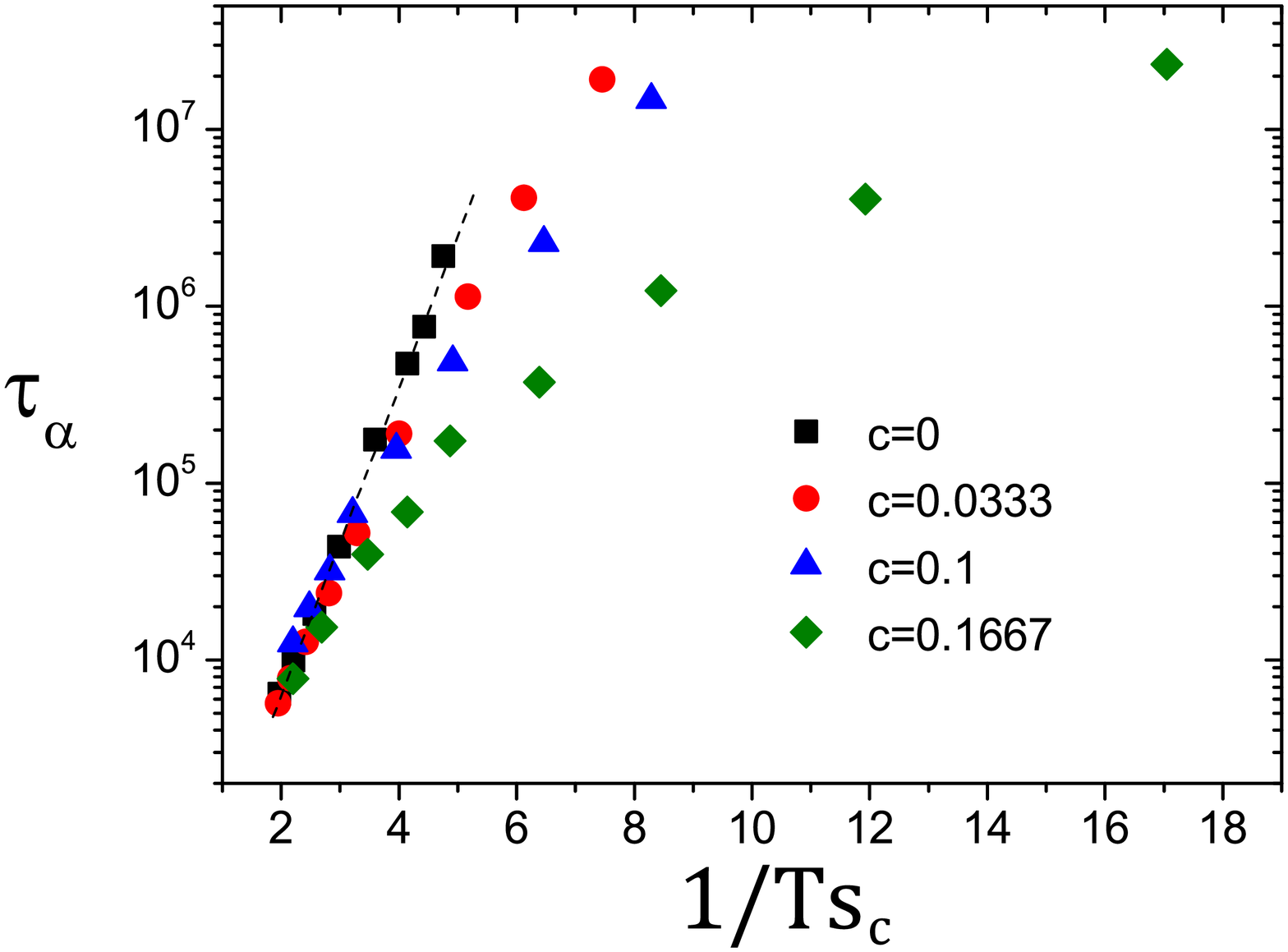}
\caption{
Logarithm of the relaxation time as a function of $1/Ts_c$. Within the
Adam-Gibbs theory one expects that this representation of the data gives
rise to a linear dependence, which for $c>0$ is evidently not the case.}
\label{fig:S2}
\end{center}
\end{figure}

We follow exactly the same method to obtain $T_d^{\rm (fit)}$ for the
pinned systems, i.e. $c\neq 0$.  In Fig.~SI-6(a), we plot $\tau_{\alpha}$
obtained from $F_s^A(k, t)$ as a function of $T-T_{d}^{\rm (fit)}$ for
several values of $c$.  Both $T_d^{\rm (fit)}$ and $\gamma$ are used
as fitting parameters to obtain the good fit with the power-law. As it
is well known for bulk systems, special care has to be taken to chose
an optimized temperature windows over which $\tau_\alpha$ can be fitted
by a power-law and we have thus put the emphasis to get a good fit in the
low temperature part of the data.

In Fig.~SI-6(b) we plot the so obtained  $T_{d}^{\rm (fit)}$ in the phase
diagram presented in the main text.  It is clear that the agreement of the
two dynamic transition temperatures obtained from different routes is very
good. However, we also recognize that the $c$-dependence of $T_{d}^{\rm
(fit)}$ is noisier.  This is most likely due to the subtleties of the
fitting procedures since three free fit parameters have to be used for
$T_{d}^{\rm (fit)}$ ($T_d^{\rm (fit)}$, $\gamma$, and prefactor), whereas
no fitting parameter is required to evaluate $T_d$ from the saddles.
This is the reason why in the present study we put more emphasis on the
static method.

\section{Violation of the Adam-Gibbs relation}

In this section we will show that for the pinned fluid
the Adam-Gibbs relation $\ln \tau_\alpha \propto 1/Ts_{\rm
c}$~\cite{SI_adam1965,SI_Biroli2012wolynesbook} is violated. In Fig.~SI-7,
we plot the relaxation time $\tau_\alpha$ against $1/Ts_c$ for different
values of $c$. For $c=0$, one observes that $\ln\tau_\alpha$ is indeed a
linear function of $1/Ts_c$ and that therefore the AG relation holds, as
it has been already documented in Refs.\cite{SI_Scala2000,SI_Sastry2001}.
However, as $c$ increases, $\log \tau_\alpha$ systematically deviates from
the linear dependence and becomes a convex function of $1/Ts_c$. Thus
this result clearly demonstrates the violation of the AG relation for
pinned fluids.

\end{document}